\documentclass[twocolumn,aps,pra,showpacs, preprintnumbers,letterpaper,amssymb,floatfix,superscriptaddress]{revtex4-2}

\usepackage[T1]{fontenc}
\usepackage{natbib}
\usepackage{graphicx}
\usepackage{bm}
\usepackage{color}
\usepackage{amsmath}
\usepackage{xspace}
\usepackage{subfigure}
\usepackage{dcolumn}
\usepackage{textcomp}
\usepackage{longtable}     
\usepackage{url}           
\usepackage[colorlinks,urlcolor=blue,citecolor=blue,linkcolor=blue]{hyperref}
\usepackage{bm}            
\usepackage{appendix}
\usepackage{hyperref}
\usepackage{soul}
\usepackage[normalem]{ulem}
\usepackage{physics}

\def\Na{$^{23}$Na\xspace}
\def\Rb{$^{87}$Rb\xspace}
\def\NaRb{$^{23}$Na$^{87}$Rb\xspace}

\begin{document}

\title{Improved characterization of Feshbach resonances and interaction potentials between $^{23}$Na and $^{87}$Rb atoms}

\author{Zhichao Guo}
\affiliation{Department of Physics, The Chinese University of Hong Kong, Hong Kong, China}
\author{Fan Jia}
\affiliation{Department of Physics, The Chinese University of Hong Kong, Hong Kong, China}
\author{Bing Zhu}
\affiliation{Hefei National Laboratory for Physical Sciences at the Microscale and Shanghai Branch, University of Science and Technology of China, Shanghai 201315, China}
\affiliation{Physikalisches Institut, Universit at Heidelberg, Im Neuenheimer Feld 226, 69120 Heidelberg, Germany}
\author{Lintao Li}
\affiliation{Department of Physics, The Chinese University of Hong Kong, Hong Kong, China}
\author{Jeremy M. Hutson}
\affiliation{Joint Quantum Centre (JQC) Durham-Newcastle, Department of Chemistry, Durham University, South Road, Durham DH1 3LE, United Kingdom}
\author{Dajun Wang}
\email{djwang@cuhk.edu.hk}
\affiliation{Department of Physics, The Chinese University of Hong Kong, Hong Kong, China}
\affiliation{The Chinese University of Hong Kong Shenzhen Research Institute, Shenzhen, China}

\date{\today}

\begin{abstract}
The ultracold mixture of \Na and \Rb atoms has become an important system for investigating physics in Bose-Bose atomic mixtures and for forming ultracold ground-state polar molecules. In this work, we provide an improved characterization of the most commonly used Feshbach resonance near 347.64 G between \Na and \Rb in their absolute ground states. We form Feshbach molecules using this resonance and measure their binding energies by dissociating them via magnetic field modulation. We use the binding energies to refine the singlet and triplet potential energy curves, using coupled-channel bound-state calculations. We then use coupled-channel scattering calculations on the resulting potentials to produce a high-precision mapping between magnetic field and scattering length. We also observe 10 additional $s$-wave Feshbach resonances for \Na and \Rb in different combinations of Zeeman sublevels of the $F = 1$ hyperfine states. Some of the resonances show 2-body inelastic decay due to spin exchange. We compare the resonance properties with coupled-channel scattering calculations that full take account of inelastic properties.


\end{abstract}

\maketitle
\section{Introduction}

\label{intro}
Feshbach resonances (FRs) provide a very flexible toolbox for controlling dilute gases of ultracold atoms~\cite{Chin2010,Kohler2006}. They allow exploration of many interesting phenomena in few-body and many-body physics~\cite{Greene2017,Bloch2008}. A Feshbach resonance occurs when a weakly bound or quasibound molecular level is shifted by an external magnetic field to be degenerate with the two-body scattering threshold. Near the resonant magnetic field $B_0$, the two-body $s$-wave scattering length for elastic scattering is approximately
\begin{equation}
a(B) = a_{\rm bg}\left(1+\frac{\Delta}{B-B_0}\right),
\label{eq1}
\end{equation}
where $a_{\rm bg}$ is the slowly varying background scattering length and $\Delta$ is the resonance width. By simply scanning the magnetic field, $a$ and thus the effective two-body contact interaction can be changed from repulsive (for $a > 0$) to attractive (for $a < 0$). Essentially arbitrary interaction strengths can be achieved. This capability has played a vital role in the study of exotic physics such as controlled collapse~\cite{donley2001}~and soliton formation in Bose-Einstein condensates (BEC)~\cite{Khaykovich2002,Strecker2002}, the BEC-BCS crossover in quantum-degenerate Fermi gases~\citep{Regal2004,Zwierlein2004,Bartenstein2004,Bourdel2004}, and more recently the formation of dilute quantum droplets~\cite{petrov2015,ferrier2016Observation,cabrera2018quantum}. In addition, FRs are used to associate atomic pairs and form weakly bound Feshbach molecules (FMs)~\cite{Kohler2006}. Combined with a subsequent two-photon Raman process, this has led to the creation of long-sought ultracold ground-state molecules~\cite{ni2008,Takekoshi2014,Molony2014,Park2015,Guo2016,Voges2020}.

In many applications, a very accurate knowledge of the mapping $a(B)$ between magnetic field and scattering length is needed. This can be provided by coupled-channel calculations on an accurate interaction potential. To obtain a suitable potential, it is necessary to fit potential parameters to high-resolution data that include accurate information on the FRs of interest.

The position of a FR is often measured by observing the peak in three-body loss that occurs near resonance. However, this method has limited accuracy, because the position and shape of the peak is influenced by complicated few-body processes. A more precise method is to measure the binding energy of FMs as function of the magnetic field~\cite{Bartenstein2005,Lompe2013}.
The binding energy can be measured spectroscopically, either by forming FMs by associating atom pairs or by dissociating FMs back to atoms. The association and dissociation process can be driven either by applying a radio-frequency field or by modulating the applied magnetic field. In general, the dissociation method offers higher precision as it is immune to thermal averaging~\cite{Bartenstein2005,Lompe2013}.

In previous work on ultracold mixtures of Na and Rb, FRs were observed by measuring 3-body losses and modeled with coupled-channel calculations \cite{Wang2013}. Binding energies of FMs were observed by magnetic-field modulation-induced association for two of the resonances, near 347.64~G and 478.7~G in collisions between \Na$\ket{F = 1, m_F = 1}$ and \Rb$\ket{F = 1, m_F = 1}$; here $F$ is the hyperfine quantum number and $m_F$ is its projection onto the axis of the field \cite{wang2015}. The resonance near 347.64~G was used to create \NaRb FMs by magnetoassociation~\cite{wang2015} and transfer them to the ground state~\cite{Guo2016}. It was also used to investigate heteronuclear few-body processes~\cite{Wang2019}. The mapping $a(B)$ between magnetic field and scattering length was obtained from the binding energies using a simplified square-well model~\cite{wang2015}. However, in recent studies of quantum droplets formed in mixtures of \Na and \Rb~\cite{guo2021}, it became clear that a more accurate mapping is needed, particularly in the vicinity of this FR.

This work reports new measurements of binding energies for the state that causes the FR near 347.64~G.
To reach the highest accuracy, we implement the dissociation method to measure the binding energies. We achieve magnetic field stability at the mG level. The data are used to refine the interaction potentials for the $X^1\Sigma^+$ and $a^3\Sigma^+$ electronic states by fitting to coupled-channel bound-state calculations. We then use coupled-channel scattering calculations to obtain a highly accurate mapping $a(B)$ for the FR near 347.64~G. This mapping provided a cornerstone for our recent experiment on the heteronuclear \Na--\Rb quantum droplet \cite{guo2021}. We also report 10 $s$-wave FRs in different combinations of Zeeman levels $m_F$ in the atomic $F = 1$ manifolds, measured by the inelastic loss method. These new resonances may find applications in future explorations of the \Na--\Rb system.

The paper is organized as follows. In Sec.~\ref{exp1}, we measure binding energies of FMs close to the FR near 347.64 G with the dissociation method. In Sec.~\ref{theor}, we refine the interaction potentials for the singlet and triplet states by fitting to the binding energies near the two resonances using coupled-channel calculations, and obtain an accurate mapping from magnetic field to scattering length. In Sec.~\ref{sec3}, we present 10 interspecies $s$-wave resonances in several different collisional channels with atoms in $F = 1$, measured by observing the enhanced atom losses. The observed resonance positions are compared with the results of coupled-channel calculations on the refined interaction potentials. In Sec.~\ref{conc}, we give the conclusions of the work.

\begin{figure}[hb]
\begin{center}
\includegraphics[width = 0.9\linewidth]{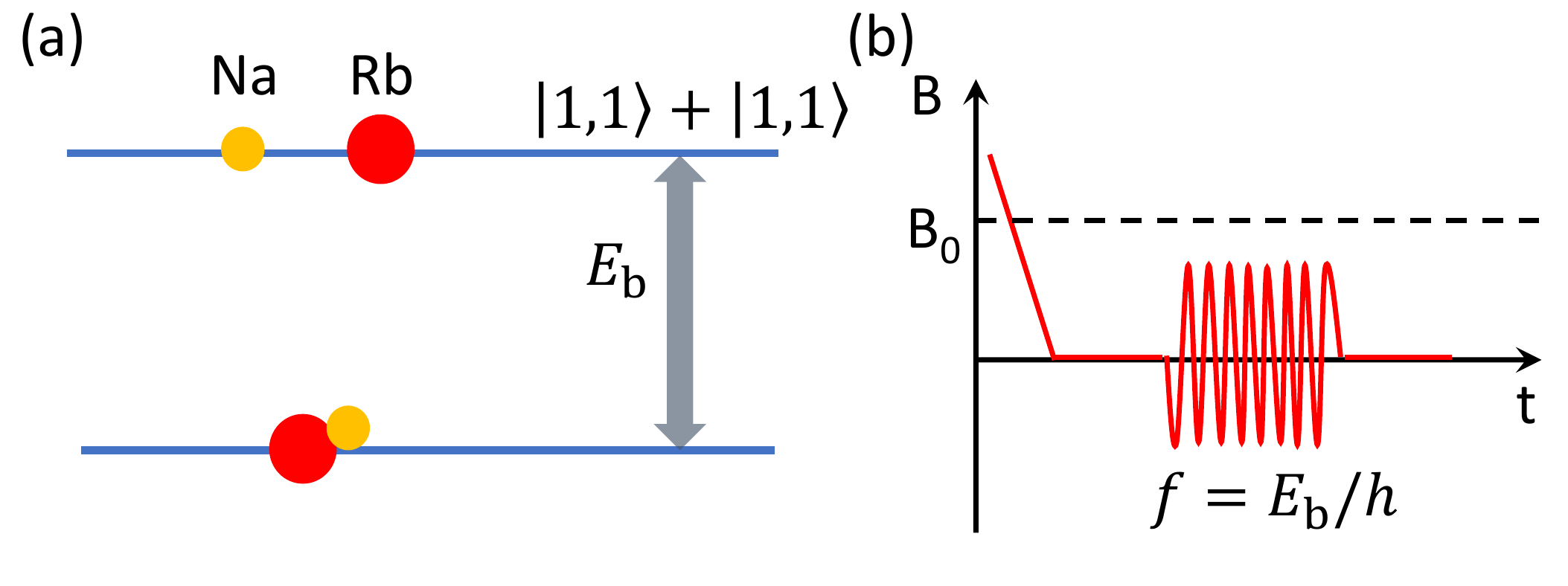}
\end{center}
\caption{Measuring the binding energy of a Feshbach molecule with magnetic field modulation spectroscopy. (a) The binding energy of the FM is $E_{\rm b}$. A bound-free transition can be driven by an oscillating magnetic field. (b) The FMs are first created by ramping the magnetic field across $B_0$. After the magnetic field is stabilized to its final value, a small-amplitude sinusoidal oscillation at frequency $f$ near $E_{\rm b}/h$ is added to dissociate the FMs and measure the binding energy. See the text for a more detailed description of the magnetic field ramping procedure.}
\label{fig1}
\end{figure}

\section{Measurement of the \NaRb Feshbach molecule binding energy}
\label{exp1}


%

The most accurate method of characterizing a FR relies on measurements of the binding energy $E_{\rm b}$ of FMs by dissociation~\cite{Bartenstein2005,Lompe2013,Chin2005Radio,Chapurin2019}.
This can be achieved by applying a radio-frequency pulse to drive a bound-free transition~\cite{Bartenstein2005,Lompe2013,Chapurin2019}. The binding energy is obtained by subtracting the free-free transition energy from the bound-free transition energy. In the current work, as shown schematically in Fig.~\ref{fig1}(a), the FM lies very close in energy to the free atom pair. In this case, the dissociation can be driven by magnetic field modulation spectroscopy~\cite{Claussen2003,Thompson2005}. As illustrated in Fig.~\ref{fig1}(b), this is implemented by adding a small-amplitude oscillation to the magnetic field after the magnetoassociation. The oscillating magnetic field can be expressed as $B + A \, {\rm sin}(2\pi f t)$, with $B$ the final magnetic field that determines $E_{\rm b}$, $A\ll B_0-B $ and $f$ the modulation amplitude and frequency, respectively. Dissociation starts to occur when $f$ matches $E_{\rm b}/h$. For $f > E_{\rm b}/h$, the excess energy is converted to kinetic energy of the free atoms as $E_{\rm k} = hf - E_{\rm b}$. Due to the variation of the bound-free Franck-Condon factor with $E_{\rm k}$~\cite{Chin2005Radio}, the dissociation spectrum is typically asymmetric and broad with respect to $f$.

Our experiment starts from an optically trapped ultracold mixture of \Na and \Rb atoms, both in their lowest hyperfine state $\ket{F = 1, m_F = 1}$~\cite{Wang2013,wang2015,Jia2020}. Magnetoassociation starts from an initial magnetic field of 350 G, just above the FR at $B_0 = 347.64$~G. The magnetic field is ramped down across the resonance to form FMs, and then to 335.6 G. At this field, the FMs have a nearly zero magnetic dipole moment; this allows us to remove the residual atoms with a short and strong magnetic field gradient without losing molecules. Afterwards, the magnetic field is ramped up to a range of target values below $B_0$ for further experiments. Following this procedure, we can routinely obtain a pure sample of \NaRb FMs with a typical temperature of 300 nK and a trap lifetime of more than 30 ms. This short lifetime is due to near-resonance photon scattering by the 947 nm optical trap light~\cite{Guo2017,Jia2020}, which is provided by a home-built diode laser system. In another experiment on \NaRb, in which a single-frequency 1064 nm laser is used as the optical trap light, FM lifetimes greater than 100 ms have been observed~\cite{Wang2019,guo2021}. Nevertheless, the current lifetime is more than enough for the present work, as we need only 10 ms for magnetic field stabilization and less than 1 ms for dissociation.

The magnetic field modulation is generated by a single-loop coil driven by a low-frequency high-power radio-frequency amplifier. The single-loop coil is placed just above the vacuum cell and coaxial with the Feshbach coils so that an oscillating magnetic field of several mG can be added to the large magnetic field. The coil has a limited modulation bandwidth of about 2 MHz. To avoid the ac Stark shift from the trapping light, the optical trap is turned off 50~$\mu$s before the modulation is applied. This also reduces possible systematic errors induced by mean-field shifts of both the FMs and the free atoms as the density of FMs is lowered. The duration of the magnetic field modulation pulse is chosen empirically so that the fraction of dissociation is no more than 70\%.

\begin{figure}[t]
\begin{center}
\includegraphics[width = 0.95\linewidth]{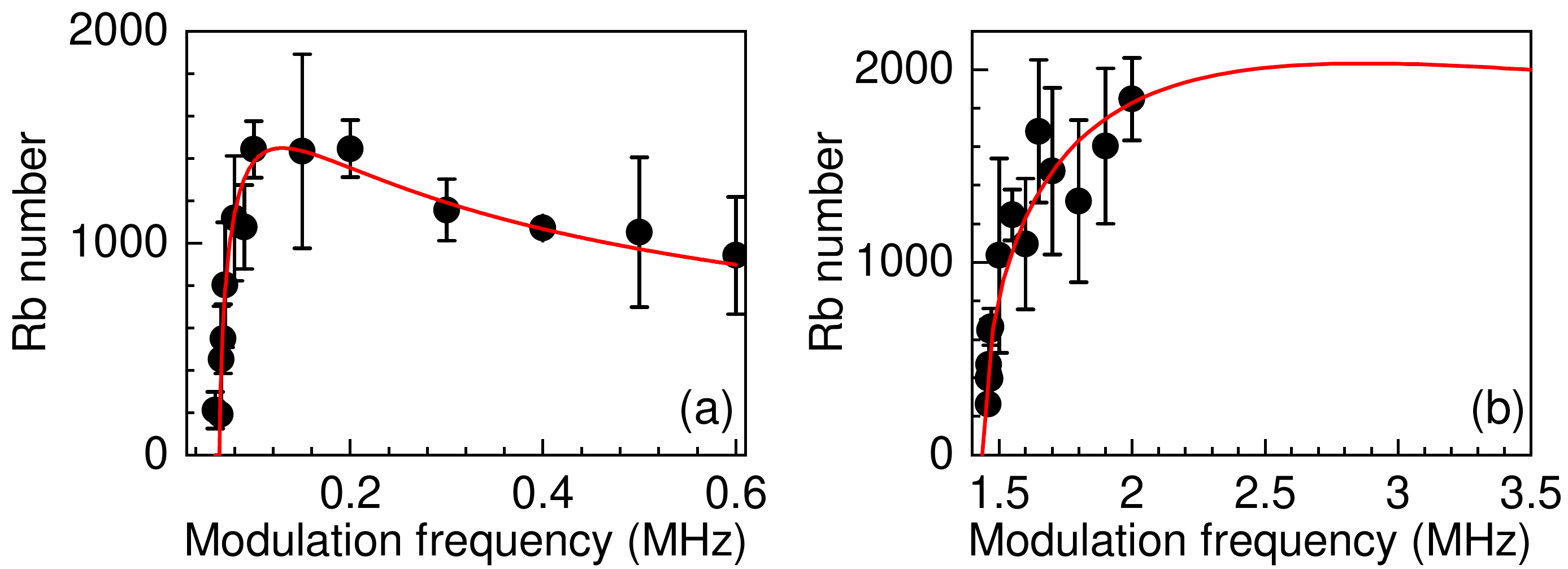}
\end{center}
\caption{Feshbach molecule dissociation spectrum at magnetic field of (a) 347.371 G, and (b) 346.000 G. The red solid curves are fitted to Eq.~\ref{eq2} to extract the FM binding energies. Because of the limited modulation bandwidth of the single-loop coil, only part of the spectrum is accessible in (b).}
\label{fig2}
\end{figure}

The dissociation signal is detected by absorption imaging of the fragment \Rb atoms. Fig.~\ref{fig2}(a) and (b) show two example dissociation spectra as a function of the modulation frequency $f$ for FMs at 347.371(5) G and 346.000(5) G, respectively. Here the magnetic field is measured with radio-frequency spectroscopy of the Rb atoms. Thresholds for dissociation are clearly visible in both spectra. For Fig.~\ref{fig2}(a), no dissociation is observed for $f$ below 60 kHz, while in Fig.~\ref{fig2}(b), the threshold is near 1.3 MHz. Since the dissociation process involves no momentum transfer, the dissociation threshold is an accurate measurement of the binding energy $E_{\rm b}$ of the FM. Above this threshold, the profile of the spectrum is determined by the overlap between the wave functions of the bound and free states. As the excess energy $hf-E_{\rm b}$ is converted into the relative motion of the two atoms, the wave function of the free atoms and thus the bound-free transition rate changes with $f$. Following~\cite{Mohapatra2015}, the lineshape of magnetic field modulation spectroscopy can be represented as
\begin{equation}
N_{\rm Rb}(f) \propto \frac{\sqrt{hf-E_{\rm b}}}{hf}.
\label{eq2}
\end{equation}
For this lineshape, a dissociation maximum occurs at $f = 2E_{\rm b}/h$; above that, the signal decays with a long tail following $1/\sqrt{f}$. The spectrum in Fig.~\ref{fig2}(a) follows this lineshape well. For the spectrum in Fig.~\ref{fig2}(b), the maximum is not reached due to the limited modulation bandwidth. The same is true for all the spectra with $E_{\rm b}/h>1$~MHz.

\begin{figure}[t]
\begin{center}
\includegraphics[width = 0.9\linewidth]{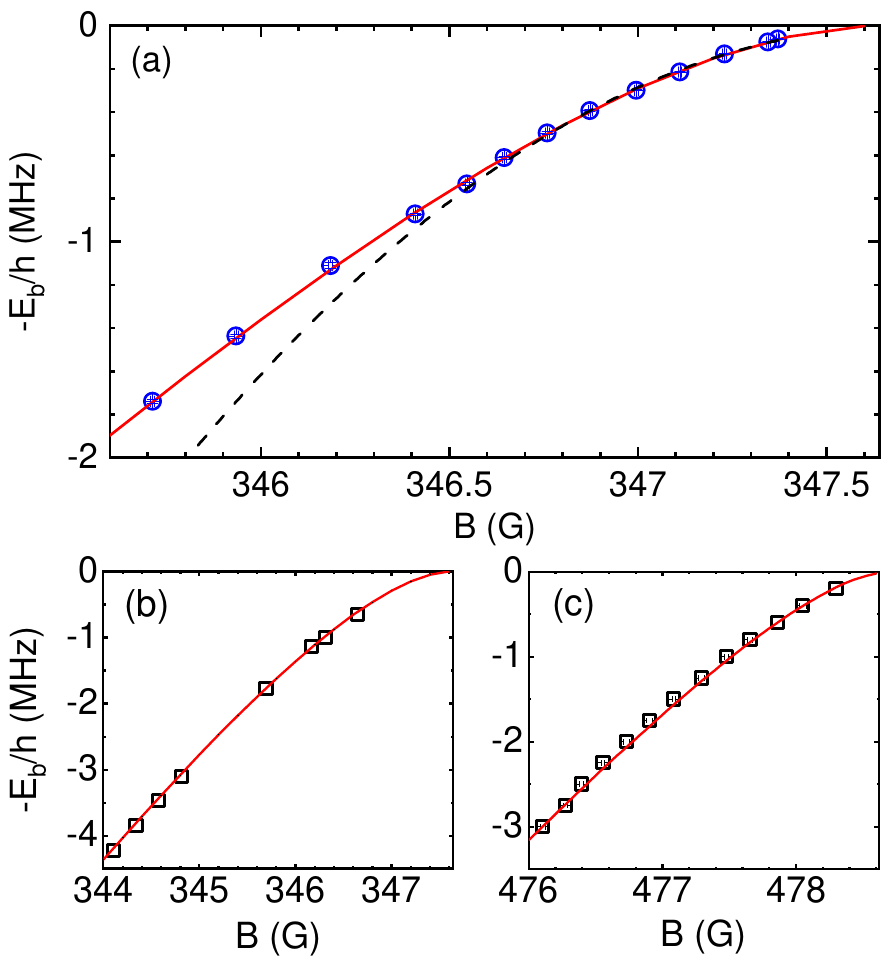}
\end{center}
\caption{(a) Binding energy of \NaRb Feshbach molecules created via the FR near 347.64 G in the entrance channel Na$\ket{1,1}+$Rb$\ket{1,1}$. The blue open circles are the data points measured in this work with the dissociation method. The error bars are smaller than the symbol size. The red solid curve is from the coupled-channel fitting, while the black dashed curve is from the universal model. (b) and (c) show the comparison between the previously observed binding energies~\cite{wang2015} (black open squares), measured using the association method, and the calculations (red solid curves) from the present coupled-channel model, for FMs created near the FRs at 347.64 G and 478.7 G, respectively.}
\label{fig3}
\end{figure}

To determine the dissociation threshold, we fit each spectrum with Eq.~\ref{eq2}. For partial dissociation spectra like Fig.~\ref{fig2}(b), we have verified with simulated data that $E_{\rm b}$ can still be determined with uncertainties less than 5 kHz. The variation of $E_{\rm b}$ with magnetic field, over a range from 0.061 MHz to 1.739 MHz, is plotted in Fig.~\ref{fig3}(a) as blue open circles. The error bar for each $E_{\rm b}$ is smaller than the symbol size. Near 347 G, the shot-to-shot fluctuation of the magnetic field, which was measured with microwave spectroscopy of $^{87}$Rb, is $\pm 3$~mG.

%
%
%

Figure~\ref{fig3} (b) and (c) show $E_{\rm b}$ for FMs that were obtained in 2015 with the association method~\cite{wang2015} near the resonances at 347.64 G and 478.7 G. In that work~\cite{wang2015}, the binding energy was fitted using the square-well model~\cite{Lange2009} with a fixed background scattering length $a_{\rm bg} = 66.77a_0$. This value of $a_{\rm bg}$ was obtained from the coupled-channel modeling of the several Feshbach resonances observed with atom-loss spectroscopy in 2013~\cite{Wang2013}. However, the square-well model requires $a_{\rm bg}$ much larger than the interaction range, which is set by the mean scattering length of the van der Waals potential \cite{Gribakin1993}; this is $\bar{a} = 55.2\,a_0$ for \NaRb. This condition is not satisfied for the Feshbach resonances considered here. It is also well known that resonance positions measured by atom-loss spectroscopy  can be affected by the complicated dynamics of three-body recombination. These issues all contribute to the inaccuracy of the Feshbach resonance parameters determined previously~\cite{wang2015}. In the following, we solve these issues with a new coupled-channel modeling of the binding energies of the FMs.

\section{Coupled-channel modeling}
\label{theor}

Coupled-channel calculations rely on expanding the total wavefunction for a pair of interacting atoms in a basis set that represents the electron and nuclear spins and the relative rotation of the atoms (the partial wave $L$). Substituting this expansion into the total Schr\"odinger equation produces a set of coupled differential equations that can be solved to obtain either bound-state or scattering properties. The Hamiltonian for the interacting pair is
\begin{equation}
\label{full_H}
\hat{H} =\frac{\hbar^2}{2\mu}\left[-\frac{1}{R}\frac{d^2}{dR^2}R
+\frac{\hat{L}^2}{R^2}\right]+\hat{H}_\textrm{A}+\hat{H}_\textrm{B}+\hat{V}(R),
\end{equation}
where $R$ is the internuclear distance, $\mu$ is the reduced mass, and $\hbar$ is the reduced Planck constant. $\hat{L}$ is the two-atom rotational angular momentum operator. The single-atom Hamiltonians $\hat{H}_i$ contain the hyperfine couplings and the Zeeman interaction with the magnetic field. The interaction operator $\hat{V}(R)$ contains the two isotropic Born-Oppenheimer potentials, for the $X^1\Sigma_g^+$ singlet and $a$ $^3\Sigma_u^+$ triplet states, and anisotropic spin-dependent couplings which arise from dipole-dipole and second-order spin-orbit coupling. In the present work, scattering calculations are carried out using the MOLSCAT package
\cite{molscat:2019,mbf-github:2020} and bound-state calculations use the related packages BOUND and FIELD \cite{bound+field:2019,mbf-github:2020}. The scattering wavefunction is expanded in a fully uncoupled basis set that contains all allowed spin and rotational functions, limited by $L_{\rm max}=2$. The numerical methods are similar to those used in Ref.\ \cite{Berninger:Cs2:2013}, so will not be described in detail here.

The interaction potential used here is based on the potential curves for the $X^1\Sigma^+$ and $a^3\Sigma^+$ states of NaRb, originally obtained by fitting to Fourier transform (FT) molecular spectroscopy~\cite{Pashov2005} and later refined by Feshbach spectroscopy with ultracold atoms~\cite{Wang2013}. The potential curve for each electronic state, with spin $S=0$ (singlet) or 1 (triplet), has three parts: (1) a high-order power-series in the well region, which is the part best determined by FT spectroscopy; (2) a long-range extrapolation, outside internuclear distance $R_{{\rm LR},S}$, which uses theoretical dispersion coefficients and a simple exchange term; (3) a short-range extrapolation, inside $R_{{\rm SR},S}$, which uses a simple repulsion of the form $A_S + B_S/R^{N_S}$. $R_{{\rm SR},S}$ is usually chosen as a distance just outside the inner turning point of the potential at zero energy. For particular values of $R_{{\rm SR},S}$ and $N_S$, $A_S$ and $B_S$ are usually determined so that the short-range extrapolation has the same value and derivative at $R_{{\rm SR},S}$ as the mid-range power-series potential \footnote{This constraint was applied in the present work, but for the potential of Ref.\ \cite{Wang2013} there are derivative discontinuities at $R_{{\rm SR},S}$.}.

Our goal is to adjust the interaction potential to fit the measured FM binding energies, while retaining as much as possible its ability to reproduce the FT spectra. We therefore keep the two power series that represent the singlet and triplet potential wells fixed at the fitted values of Ref.\ \cite{Wang2013}. We also retain the long-range extrapolation in its original form, with the dispersion coefficients unchanged. We vary only the parameters $R_{{\rm SR},S}$ and $N_S$ that define the short-range extrapolations. For the triplet potential, $R_{{\rm SR},1}$ is held at its original value from Ref.\ \cite{Wang2013}, and $N_1$ is varied; this provides sufficient flexibility to adjust the singlet scattering length and reproduce resonance positions and binding energies. For the singlet potential, however, the original value of $R_{{\rm SR},0}$ is so close to the turning point that varying $N_0$ does not allow enough change in the potential to reproduce the experimental data. In this case $N_0$ is held at its original value from Ref.\ \cite{Wang2013}, and $R_{{\rm SR},0}$ is varied.

The shape of the curve of $E_{\rm b}$ as a function of $B$ is quite insensitive to variations in the singlet and triplet scattering lengths, represented by $R_{{\rm SR},0}$ and $N_1$. It is therefore sufficient to fit to one binding energy from each FR. For the resonance near 347.64 G, we choose the point with $E_{\rm b}/h = 0.611(6)$~MHz at 346.646(5) G, while for the resonance near 478.7 G, we choose the point with $E_{\rm b}/h = 0.400$~MHz at 478.052(10) G. The parameters $R_{{\rm SR},0}$ and $N_1$ are then adjusted so that these binding energies are reproduced essentially exactly. The resulting values are $R_{{\rm SR},0}=2.7256$~\AA\ and $N_1=11.492$. 

The fitted potentials are used to calculate the binding energies of the bound states that cause the resonances near 347.64 G and 478.7 G as a function of $B$, using coupled-channel calculations as described above. We also calculate the singlet fraction and the magnetic moment with respect to free atoms. The calculated binding energies $E_{\rm b}^{\rm cc}$ are shown in Fig.~\ref{fig3} as solid red lines; they agree with the experimental data very well, even though only 2 data points are used in the fitting. The relative residuals $E_{\rm b}/E_{\rm b}^{\rm cc}-1$ are all within $\pm 2\%$.

We use the fitted potentials to calculate the $s$-wave scattering length $a(B)$. We find that each resonance is well represented by Eq.~\ref{eq1}. The parameters $B_{0}$, $\Delta$ and $a_{\rm bg}$ are obtained as described by Frye and Hutson \cite{Frye2017}, and are given in Table~\ref{tab1}. In Fig.~\ref{fig3}(a), the binding energies calculated from the universal model $E_{\rm b} = \hbar^2/2\mu a(B)^2$ using the new $a(B)$ is also shown. Compared with the coupled channel results (red solid curve), this reproduces the data only for $E_{\rm b}< 0.6$ MHz.

\begin{table}[tb]
\caption{Parameters of the two low-field $s$-wave Feshbach resonances for the entrance channel Na$\ket{1,1}$+Rb$\ket{1,1}$.}
\label{tab1}\centering
\begin{tabular}{ c | c | c | c }
\hline\hline
$B_0$ (G) & $\Delta$ (G) & $a_{\rm bg}$ ($a_0$) & reference\\
\hline
347.648(33) & 4.259(2) & 76.33(1) & this work\\
347.64 & 5.20  & 66.77 & \cite{wang2015}\\
347.75 & 4.89  & 66.77 & \cite{Wang2013}\\
\hline
478.714(28) & 3.495(3) & 71.48(7) & this work\\
478.83 & 4.81  & 66.77 & \cite{wang2015}\\
478.79 & 3.80  & 66.77 & \cite{Wang2013}\\
\hline \hline
\end{tabular}
\end{table}

A complete error analysis would require refitting in the full parameter space of ref.\ \cite{Wang2013}, which has more than 50 dimensions, and including all the experimental data from FT spectroscopy. This is beyond the scope of the present work, and indeed has not been done for any of the alkali-alkali interaction potentials fitted to FT spectroscopy by the Hannover group. In this work we focus on the parts of the potential needed to give a good account of the magnetic Feshbach resonances near 347.64 and 478.7 G. These properties are principally sensitive to the singlet and triplet scattering lengths, which are adjusted using $R_{\textrm{SR},0}$ and $N_1$. Nevertheless, to estimate the way in which uncertainties in magnetic fields and binding energies propagate, we have repeated the fits with the two data points described above shifted in either direction by their 1$\sigma$ uncertainties. This yields potential parameters $R_{{\rm SR},0}=2.7256(6)$~\AA\ and $N_1=11.492(100)$. The uncertainties in resonance parameters resulting from this are included in Table I. It should be noted that there are additional uncertainties due to model dependence, and to parameters not refitted here. It is not practical to evaluate these in a systematic way.

The singlet and triplet scattering lengths calculated from the fitted potentials are $a_{\rm s} = 106.487(3)\,a_0$ and $a_{\rm t} = 68.859(6)\,a_0$, respectively. These are $0.27\,a_0$ smaller and $0.24\,a_0$ larger than the values of Ref.\ \cite{Wang2013}, indicating that the triplet potential needs to be slightly more repulsive than the original at short range, while the singlet potential needs to be slightly less repulsive.

\begin{figure}[ht]
\begin{center}
\includegraphics[width = 0.9\linewidth]{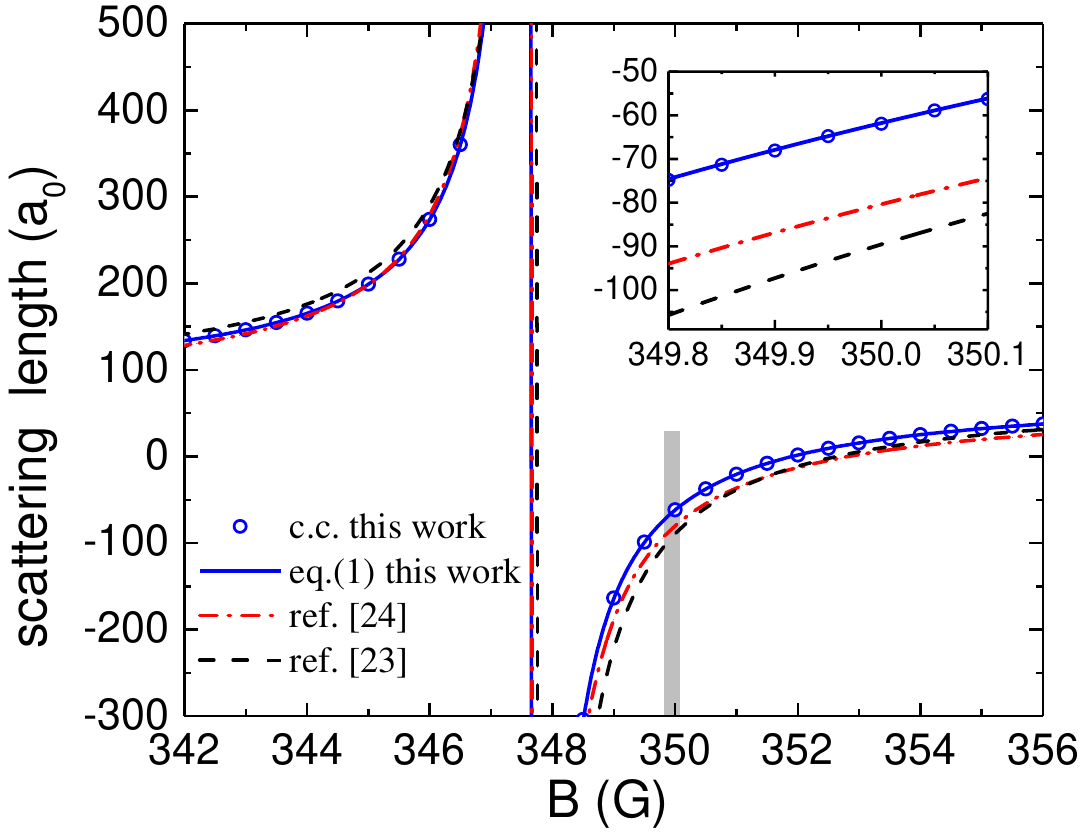}
\end{center}
\caption{Comparison of the mapping between magnetic field and scattering length in unit of Bohr radius ($a_0$)  with Feshbach resonance parameters obtained in this work, using coupled-channel results directly (blue open circles) and fitted to Eq.~\ref{eq1} (blue solid curve) and , and in Ref.~\cite{wang2015} (red dash-dotted curve) and Ref.~\cite{Wang2013} (black dashed curve), both using Eq.~\ref{eq:FRsum}. The region of interest for the droplet experiment~\cite{guo2021}, marked with a vertical gray bar, is enlarged in the inset. }
\label{fig_asB}
\end{figure}

In the present fit, the resonance near 347.64 G has shifted by about 0.08 G compared to Ref.\ \cite{Wang2013}. The resonance near 487.7 G has shifted by more than 0.1 G. These shifts produce substantial changes in the calculated scattering lengths. Moreover, in Ref.\ \cite{Wang2013}, the calculated scattering length was fitted using the formula
\begin{equation}
a(B)=a_{\rm bg} \left[1 + \frac{\Delta_0}{B-B_0} + \frac{\Delta_1}{B-B_1} + \dots\right],
\label{eq:FRsum}
\end{equation}
with a single $B$-independent background scattering length $a_{\rm bg}$ for all resonances. This formula breaks down when the background scattering length varies significantly between resonances. In the present work, Eq.~\ref{eq1} is found to be a good local representation for each resonance, but significantly different values of $a_{\rm bg}$ are needed for the resonances at 347.64 G and 487.7 G. Such behavior can arise from many sources, including the changing spin character of atomic states with magnetic field and the presence of additional resonances not included in the summation of Eq.\ \ref{eq:FRsum}. As shown in Fig.~\ref{fig_asB}, for the resonance near 347.64 G, the scattering lengths calculated with the new potential differ by about $20a_0$ from those of Ref.\ \cite{wang2015} in the region from 349.8 G to 350.0 G that is important for studies of Na-Rb quantum droplets. This change is significant and largely explains the discrepancies initially observed between experiment and theory in Ref.\ \cite{guo2021}.

All the coupled-channel calculations use a basis set that includes functions for partial waves $L = 0$ and $L =2$, with the effective dipolar coupling function of Ref.\ \cite{Wang2013} unchanged. Omitting the $L = 2$ basis functions causes shifts that are on the same level as the experimental uncertainties: the calculated resonance positions and binding-energy curves shift down by between 3 and 4 mG. It is possible to include additional partial waves, but we expect the shifts to be very small. In contrast to Ref.\ \cite{Wang2013}, we do not include any variation of the atomic hyperfine coupling with internuclear distance.

\section{Additional Feshbach resonances}
\label{sec3}

So far, most experimental work on the \Na-\Rb system has been performed with the FR near 347.64 G. However, \Na-\Rb actually has very rich resonance structure in collisions between atoms in $F=1$ states. In this section, we investigate that structure. For simplicity, we denote the state of a pair of \Na + \Rb atoms by $\ket{m_F^{\rm Na}}+\ket{m_F^{\rm Rb}}$. Ref.~\cite{Wang2013} investigated the entrance channels $\ket{1}+\ket{1}$ and $\ket{-1}+\ket{-1}$ and in total observed three $s$-wave and two $p$-wave FRs. In addition, ten more $s$-wave FRs were predicted for collisions between pairs of atoms in different $F = 1$ hyperfine Zeeman states. In this section, we report the experimental observation of several of these FRs below 1000 G in 5 hyperfine Zeeman combinations. We believe these FRs will find important applications in the future, for example in the investigation of BEC mixtures and quantum droplets with more than two components~\cite{ma2021}.

\begin{figure*}[tb]
\begin{center}
\includegraphics[width = 0.9 \linewidth]{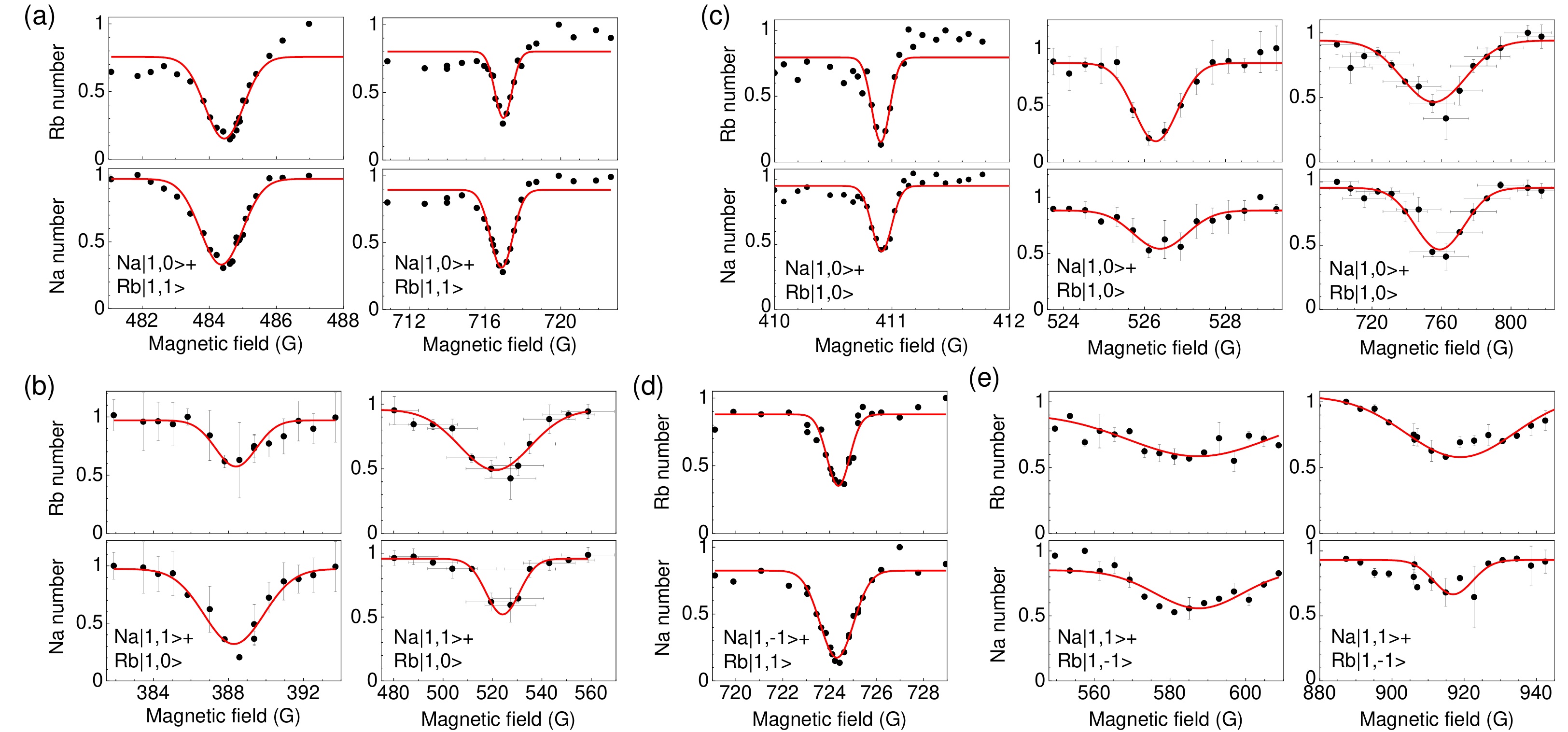}
\end{center}
\caption{Feshbach resonances in \Na-\Rb, observed by losses for atoms in the $F = 1$ hyperfine states. (a) and (b) are for the two channels in the $M_F = 1$ manifold; (c), (d) and (e) are for the three channels in the $M_F = 0$ manifold. The typical holding time at each magnetic field is 50 ms. The solid curves are from Gaussian fitting to determine the lineshape center. Error bars for the atom numbers represent one standard deviation. The large error bars for the magnetic field for the several broad loss spectra are due to less precise magnetic field control in these cases (see text).}
\label{fig4}
\end{figure*}

\subsection{Experimental observation with atom loss spectroscopy}

The experiment for this section starts from optically trapped thermal mixtures of \Na and \Rb atoms in the states $\ket{-1}+\ket{-1}$. The number of atoms for each species is typically around $10^5$ and the sample temperatures are about 1~$\mu$K. The atoms are then transferred to selected hyperfine Zeeman levels by radio-frequency rapid adiabatic passage. For each species, the Zeeman splittings $\ket{-1}\rightarrow \ket{0}$ and $\ket{0}\rightarrow \ket{1}$ are very similar. In addition, these splittings in different species are very similar to each other. To avoid cascade transitions and to realize species-selective state control, the state transfers are performed at 100 G, where the transition frequencies all differ by more than 500 kHz. With this method, we are able to prepare the \Na--\Rb mixture in all the 9 possible $m_F$ combinations of the $F = 1$ states.

After the state preparation, we ramp the magnetic field to different values and hold the samples for typically 50 ms before releasing the atoms from the optical trap and measuring the numbers of atoms remaining. Interspecies FRs manifest as losses of both \Na and \Rb atoms, as a result of enhanced interspecies three-body recombination rates or enhanced 2-body inelastic processes. As shown in Fig.~\ref{fig4}, we observe 10 FRs in 6 hyperfine Zeeman combinations.

\begin{figure}[tb]
\begin{center}
\includegraphics[width = 0.95 \linewidth]{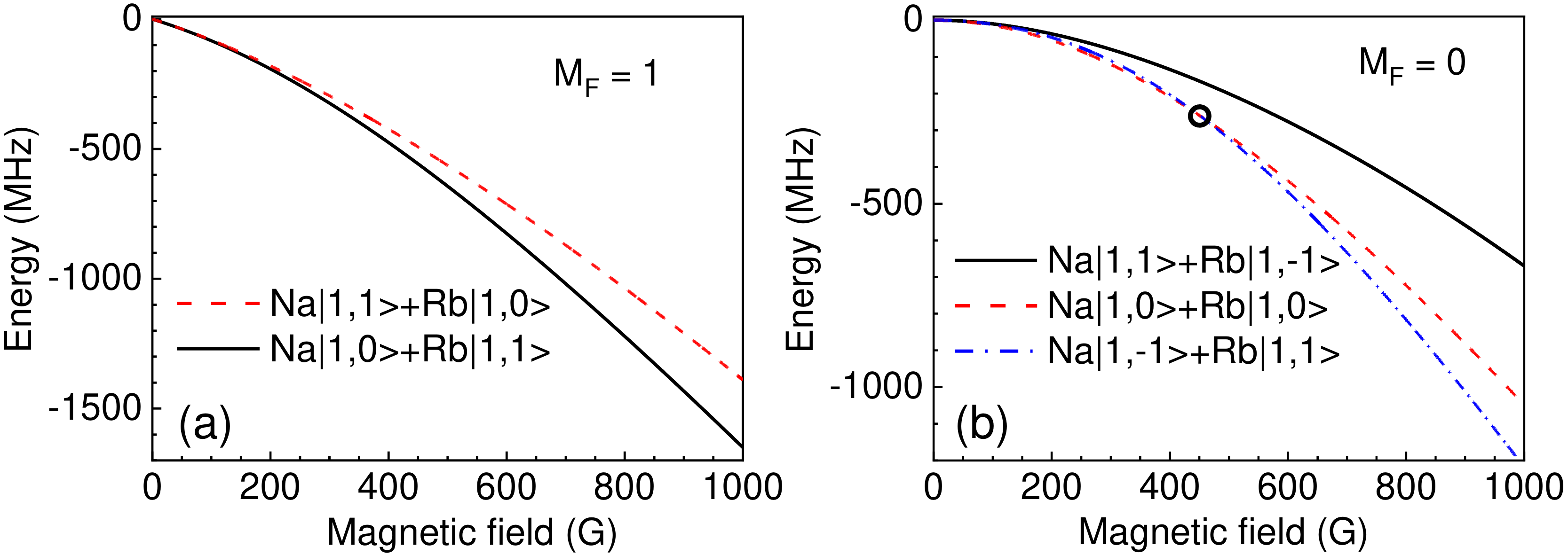}
\end{center}
\caption{(a) Channel energies for the manifold with $M_F = 1$ for magnetic field up to 1000 G. (b) The same for the manifold with $M_F = 0$. The energies of the channels $\ket{0}+\ket{0}$ and $\ket{-1}+\ket{1}$ cross each other near 450 G, as marked by the black open circle. For each $M_F$, atom pairs in higher-energy channels can undergo inelastic loss by spin exchange to the lower-energy channels.}
\label{fig5}
\end{figure}

The observed loss features have a wide range of widths. This is partly due to the availability of different inelastic loss channels for different entrance channels. For two-body collisions in a central field, the total angular momentum projection $M_F = m_F^{\rm Rb} + m_F^{\rm Na}$ is conserved. However, when $|M_F|<2$ there is more than one channel with very similar energy. Fig.~\ref{fig5}(a) shows the threshold energies of the two channels with $M_F = 1$ in magnetic field up to 1000 G. The $\ket{1} + \ket{0}$ always has lower energy than $\ket{0} + \ket{1}$. Correspondingly, the two loss features for the entrance channel $\ket{0} + \ket{1}$, shown in Fig.~\ref{fig4}(b), are broadened by inelastic loss due to spin exchange to the channel $\ket{1} + \ket{0}$~\cite{Li2015,Li2020}. The three channels with $M_F = 0$, shown in Fig.~\ref{fig5}(b), are more complicated.
The lowest-energy channel is $\ket{-1}+\ket{1}$ for magnetic fields below 1 G~\cite{Li2020}, $\ket{0}+\ket{0}$ between 1 G and 450 G, and $\ket{-1}+\ket{1}$ above 450 G. The highest-energy channel is always $\ket{1}+\ket{-1}$.
As shown in Fig.~\ref{fig4}(c), for the entrance channel $\ket{0}+\ket{0}$, the resonant feature near 411 G is very narrow, while the other two features near 526 G and 760 G are increasingly broad. For the entrance channel $\ket{-1}+\ket{1}$, only a narrow resonance near 724 G is observed. For the entrance channel $\ket{1}+\ket{-1}$, all the observed loss features are very broad.

Experimentally, for the loss spectra with widths narrower than 10 G, the magnetic field is controlled by a high-stability feedback system and also calibrated locally with rf spectroscopy on \Rb atoms immediately after taking each loss spectrum. The measured magnetic fluctuation is around 10 mG. For very broad spectra, the magnetic field is controlled less precisely and the fluctuation is larger.

To extract the resonance center for each resonance, we fit the loss spectra of both the \Rb and \Na atoms to Gaussian functions, giving the red curves in Fig.~\ref{fig4}. For all the resonances, the parameters obtained from the fitting are not the same for the two species. The resonance centers $B_0^{\rm exp}$ listed in Table~\ref{fst} are the means of the fitted resonance centers of \Rb and \Na. The center and width measured in the loss spectroscopy are likely to be shifted from the $B_0$ and $\Delta$ parameters of a FR as in Eq.~\ref{eq1} because of the complicated 2-body and 3-body physics of loss.

\subsection{Theoretical modeling}

\begin{table*}[th]
\caption{Comparison of experimental resonance positions $B_0^{\rm exp}$ with theoretical parameters for interspecies FRs below 1000 G in \Na--\Rb, for the nine $F = 1$ entrance channels. $L$ indicates the partial wave of the entrance channel. $B_0^{\rm exp}$ is the mean of the centers of the loss spectra for \Na\ and \Rb in Fig.~\ref{fig4}, determined by Gaussian fitting. Error bars represent one standard deviation. The theoretical values are from coupled-channel calculations using the singlet and triplet potential-energy curves described in Sec.~\ref{theor} with the latest potential parameters. $B_0^{\rm cc}$, $\Delta$, and $a_{\rm bg}$ are the theoretical position, elastic width, and background scattering length, respectively. The last column ``inel.?'' indicates whether the FR is subject to inelastic losses from spin exchange. For decayed FRs with inelastic losses, the resonant scattering length $a_{\rm res}$ and the inelastic width $\Gamma_{\rm inel}$ are also listed.}
\label{fst}\centering

\begin{tabular}{l|c|l|c|c|c|c|c|c}
\hline\hline
Entrance channel 				& $L$ & $B_0^{\rm exp}$(G)	& $B_0^{\rm cc}$(G)	& $\Delta$(G)	& $a_{\rm bg}$($a_0$) & $a_{\rm res}$($a_0$) & $\Gamma_{\rm inel}$(G)	& inel.? \\
\hline
Na$\ket{1,1}$ + Rb$\ket{1,1}$   & 1 	& 284.1(3)        	& 283.894			&   						 &   		& & & N     \\
						    	& 1    	& 284.2(3)         	& 283.936 			& 	&       							 & & & N     \\
    							& 1    	& 284.9(3)         	& 284.735 			&   &         							 & & & N     \\
    							& 1    	& 285.1(3)          & 284.993 			&   &         							 & & & N     \\
								& 0    	& 347.61(2)       	& 347.645 			& 4.258  			 & 76.328  	& & & N     \\
					        	& 0 	& 478.82(3)        	& 478.712 			& 3.495  			 & 71.548   & & & N     \\ \hline
Na$\ket{1,1}$ + Rb$\ket{1,0}$   & 0    	& 388.5(2)          & 388.577 			& 5.684  & 78.441                 &6548.8 & $-0.13617$& Y     \\
								& 0    	& 522(10)           & 524.286 			& 1.010  & 74.332                 & 12.906& $-11.634$& Y     \\ \hline
Na$\ket{1,0}$ + Rb$\ket{1,1}$   & 0   	&  	--		        & 358.078 			& <0.001  & 80.108                & & & N     \\
								& 0   	& 484.45(5)         & 484.569 			& 4.476  & 77.258                 & & & N     \\
								& 0		& 	--				& 570.441			& <0.001	& 73.784			 & & & N \\
								& 0   	& 716.97(6)         & 717.166 			& 5.593  & 72.998                 & & & N     \\  \hline
Na$\ket{1,1}$ + Rb$\ket{1,-1}$ 	& 0    	& 587(3)         & 580.686 			& 0.241  & $82.389 - 0.0072 i $  & $2.3917 + 0.87566 i$&$-16.596$& Y     \\
							  	& 0    	& 920(1)         & 913.431 			& 0.091  & $81.139 + 0.0352 i$ & $0.47768 + 0.12885 i$& $-30.861$& Y     \\ \hline
Na$\ket{1,0}$ + Rb$\ket{1,0}$  	& 0    	& 410.90(1)         & 409.810 			& 0.176  & 82.328                 & & & N     \\
    							& 0    	& 526.29(3)         & 526.548 			& 5.676  & 78.132                 &23515 &$-0.03772$ & Y     \\
    							& 0    	& 759(13)           & 768.781 			& 1.638  & 75.495                 &19.983 &$-12.373$ & Y     \\ \hline
Na$\ket{1,-1}$ + Rb$\ket{1,1}$ 	& 0    	& --		        & 419.648 			& <0.001  & 80.109                &0.25533 &$-0.34477$ & Y     \\
								& 0    	& --        		& 516.986 			& 0.003  & 79.401                 & & & N     \\	
								& 0    	& 724.38(4)         & 727.270 			& 4.870  & 76.444                 & & & N     \\ \hline
Na$\ket{1,0}$ + Rb$\ket{1,-1}$  & 0    	& --          		& -- 			& -- & --                             & & & N     \\ \hline
Na$\ket{1,-1}$ + Rb$\ket{1,0}$  & 0    	& --          		& 609.754 			& 0.416 & 80.572                  & & & N     \\
								& 0		& --				& 759.441			& 5.676 & 78.826				  & & & N \\ \hline	
Na$\ket{1,-1}$ + Rb$\ket{1,-1}$ & 0    	& 899.8(3)          & 900.317 			& -0.315 & 79.270                 & & & N     \\
								& 1    	& 954.2(3)          & 954.755 			&   &         				      & & & N     \\
								& 1    	& 954.5(3)          & 955.024 			&   &         					  & & & N     \\
\hline \hline
\end{tabular}

\end{table*}

We have carried out coupled-channel calculations of resonance parameters for both elastic and decayed resonances, using the interaction potentials obtained in Sec.~\ref{theor}. The results are included in Table~\ref{fst}. For elastic resonances, the resonance positions for the lowest threshold of each $M_F$ is first located with the FIELD program~\cite{bound+field:2019,mbf-github:2020}. The FRs are then characterized using the MOLSCAT program~\cite{molscat:2019,mbf-github:2020} following the elastic procedure of Ref.~\cite{Frye2017}, which converges on the resonance position $B_0^{\rm cc}$, the elastic resonance width $\Delta$, and the background scattering length $a_{\rm bg}$. Here $a_{\rm bg}$ is a local background scattering length that is different for each resonances. In the current calculation, we have not included the spin-spin interaction, so that both the quantum number $L$ of the partial wave and its projection $M_L$ are conserved in the calculation. Since $M_{\rm tot} = m_F^{\rm Na}+m_F^{\rm Rb} + M_L$ is conserved, only closed channels with $M_F = m_F^{\rm Na}+m_F^{\rm Rb}$ the same as the colliding atoms can cause FRs.

When an inelastic channel is present, the states at threshold are quasi-bound, so using FIELD is more complicated. Instead, the FRs are located by computing the scattering lengths for magnetic fields up to 1000 G using the MOLSCAT program~\cite{molscat:2019,mbf-github:2020}. Once a FR is located, the resonance parameters are then determined following the regularized scattering length or fully complex procedure of Ref.~\cite{Frye2017}. In this case, the scattering length is complex and shows an oscillation rather than a pole at resonance \cite{Hutson:res:2007}. Such a resonance is termed decayed, and the procedure generates the resonant scattering length $a_{\rm res}$ and the inelastic width $\Gamma_{\rm inel}$ in addition to $B_0^{\rm cc}$, $\Delta$ and $a_{\rm bg}$.

The calculation reproduces all the observed $s$-wave resonances. As shown in Table~\ref{fst}, the deviations between $B_0^{\rm cc}$ and $B_0^{\rm exp}$ are within 0.5 G for most of the elastic FRs, labeled by ``N'' in the last column ``inel?''. The only exception is the resonance observed at 410.9 G for the entrance channel $\ket{0} + \ket{0}$, for which $B_0^{\rm cc}$ is 1.1 G lower. As shown in Fig.~\ref{fig5}(b), this channel features a nearby switchover between the relative channel energies near 450 G. The four $s$-wave resonances for the entrance channels $\ket{1}+\ket{1}$ and $\ket{0}+ \ket{1}$ all have calculated resonance widths $\Delta$ of several Gauss. They should all be useful for investigating \Na--\Rb mixtures with tunable interactions. These resonances also have rather small calculated effective ranges, on the order of 10s of $a_0$; thus for most applications, using the scattering lengths should be sufficient.

The decayed FRs can have very different characters, depending on $a_\textrm{res}$ and/or $\Gamma_\textrm{inel}$. When $a_\textrm{res}$ is large and $\Gamma_\textrm{inel}\ll\Delta$, the resonance \emph{may} be fairly similar to the elastic case and be dominated by three-body loss. However, when $a_\textrm{res}$ is small and $\Gamma_\textrm{inel}$ is comparable to or larger than $\Delta$, the loss is more likely to be dominated by two-body loss. This looks to be the case for the resonances at 522 G for $\ket{1}+\ket{0}$, at 759 G for $\ket{0}+\ket{0}$, and the unobserved one at 420 G for $\ket{-1}+\ket{1}$. It is also true for both resonances for $\ket{1}+\ket{-1}$, but for those there is a significant \emph{background} loss characterized by the imaginary part of $a_\textrm{bg}$.

For the decayed resonances with $\Gamma_\textrm{inel} <1$ at 388 G for $\ket{1}+\ket{0}$ and at 526 G for $\ket{0}+\ket{0}$, the calculated resonance positions $B_0^{\rm cc}$ agree with the measured resonance positions $B_0^{\rm exp}$ very well. However, deviations up to several Gauss are observed for inelastic resonances with large $\Gamma_\textrm{inel}$. These can be attributed to several factors. For the two resonances with very broad observed loss spectra at 522 G for $\ket{1}+\ket{0}$ and at 759 G for $\ket{0}+\ket{0}$, the resonance centers have large uncertainties near 10 G. In addition, in presence of the background loss, the two-body loss induced by these resonances has an asymmetric (Fano-like) profile and its measured peak is not at $B_0$. Finally, although we have not studied them carefully, some nearby intraspecies FRs in Na or Rb may also affect the measurements of the resonance positions.

Several resonances predicted by the coupled-channel calculation are not observed experimentally. Most of these are very narrow ones with calculated $\Delta$ in the mG range. No resonance is found from the calculation for the entrance channel $\ket{0}+\ket{-1}$. In addition, we have not searched experimentally for FRs in the entrance channel $\ket{-1}+\ket{0}$ although the calculation indicates that two resonances should exist.

Table~\ref{fst} also lists the several $p$-wave FRs observed by~\cite{Wang2013} for the entrance channels $\ket{1}+\ket{1}$ and $\ket{-1}+\ket{-1}$. As mentioned above, we have not attempted to improve the agreement between the current calculation and the experiment by including the dependence of the atomic hyperfine coupling on $R$. We note that Ref.~\cite{Cui2018}, which presented a compilation of FRs for alkali atoms calculated with multichannel quantum defect theory (MQDT), identified the resonance near 284 G as a ``broad'' $p$-wave resonance.

It would be possible to perform a full-scale coupled-channel fitting to these resonances. This, however, is not pursued here as it would involve further adjustments to the potential parameters, beyond those used in Ref.\cite{guo2021}. In view of the inaccuracy of the loss spectroscopy compared with the measurements of FM binding energies, we believe such fitting is not currently justified.

\section{Conclusion}
\label{conc}

We have measured precise binding energies for the bound state responsible for the broad Feshbach resonance near 347.64 G in the lowest Zeeman hyperfine channel of \Na-\Rb. The measurements use magnetic field modulation spectroscopy to dissociate Feshbach molecules. We have used these measurements, together with earlier measurements \cite{wang2015} of the bound states responsible for the resonance near 478.7 G, to refine the repulsive part of the potential-energy curves for the singlet and triplet electronic states, using coupled-channel bound-state calculations. We have then used coupled-channel scattering calculations on the refined interaction potentials to provide a precise mapping $a(B)$ from magnetic field to scattering length. The mapping differs substantially from earlier results \cite{Wang2013,wang2015}. It sets the stage for experiments on ultracold mixtures of \Na--\Rb with tunable interactions, such as quantum droplet formation and stability~\cite{guo2021}, polaron physics, and Efimov physics~\cite{Wang2019}.

We have also measured 10 Feshbach resonances in a variety of Zeeman combinations of \Na($F=1$) and \Rb($F=1$). We have carried out coupled-channel scattering calculations of the resonance parameters, including positions and both elastic and inelastic widths. For resonances where only elastic scattering is possible, or inelastic decay is weak, the positions are in generally good agreement with experiment. However, there are other resonances for which the calculations identify strong inelastic decay, which can shift or broaden the resonances and in some cases makes then unobservable by the present methods. The rich resonance structure in the \Na--\Rb\ system makes it a candidate for studying multiple-component ultracold mixtures with two tunable interspecies interactions~\cite{ma2021}. In addition, there are resonances with partial-wave quantum number $L=1$ that may be useful for investigating quantum gas mixtures with non-zero relative angular momentum~\cite{Cui2017,Cui2018}.

\section{Acknowledgments}
We thank Bo Gao, Paul Julienne, Gaoren Wang, and Yue Cui for valuable discussions. This work was supported by the Hong Kong RGC General Research Fund (grants 14303317 and 14301818) and the Collaborative Research Fund C6005-17G. JMH is supported by the U.K.\ Engineering and Physical Sciences Research Council (EPSRC) Grant No.\ EP/P01058X/1.


%

\end{document}